\documentstyle[11pt]{article}
\begin{document}
\begin{center}                    
{\bf  Analytical Description of Classical Trajectories and Deflection Function in  
Scattering}\\
\vspace{0.5cm}
S. K. Gupta, Arun K. Jain and B. M. Jyrwa$^{*}$,

Nuclear Physics Division,\\
Bhabha Atomic Research Centre,\\
Mumbai 400 085, India\\
  
$^{*}$Physics Department,
North-Eastern Hill University,\\
Shillong 793 022, India 
\end{center}
\vspace{0.5cm}
\hspace{1cm}
\begin{abstract}
Analytical expressions are derived for classical trajectories in repulsive
Coulomb plus multi-step attractive potentials. Thereafter
the closed form expressions are obtained for the classical deflection functions.
The expressions are expected to be of use in heavy ion interactions in nuclear
physics.

\end{abstract}

\section{Introduction}

 Ford and Wheeler [1] demonstrated  that under the conditions of applicability
of semiclassical analysis of quantal scattering the quantum-mechanical scattering
amplitude can be simply related to the classical deflection function.
Many of the interesting characteristics of  scattering  are related to various 
features of the classical deflection function. In literature usually the 
closed-form expression for the deflection function is available only for the Coulomb 
potential see e.g.[1-4]. In  this paper, the closed-form results for attractive staircase 
 potential having several steps plus the repulsive Coulomb potential 
are being presented. Though  the formalism described here is general and is 
applicable to all cases of scattering, it is especially suited to describe
heavy ion scattering in nuclear physics [3,4]. We also give expressions for the
classical trajectories because many classical and semiclassical formulations  use the trajectories 
in the description of scattering  and other phenomena such as fusion or deep inelastic  
scattering.

\section{ Trajectories and deflection function}

In the centre of mass system, the scattering between two charged particles reduces to
that of a point particle at energy $E$ being scattered by a central potential $V(r)$, 
where $r$  is the 
inter-particle distance. For a given impact parameter $b$, the trajectory is  
reflection-symmetric about the bisector of the angle between the initial and 
final directions. The bisector intersects the trajectory at a distance $ r = a $,  
where $a$ is the distance of closest approach measured from the origin.  Here
the angle $\phi$  is measured with respect to the  bisector. For impact parameter $b$,
the trajectory, i.e. a relationship between  angle $ \phi$ and $r$ is given by 

\begin{equation} 
\phi(b,r)=\int_{a}^{r} dr \frac{ b}{ r^2 \sqrt{1-\frac{V(r)}{E}-\frac{b^2}{r^2}}}.
\end{equation} 

where $a$ is  given by the solution of the following equation
\begin{eqnarray} 
a^2-\frac{V(a)}{E}a^2-b^2 = 0.
\end{eqnarray}

The deflection function, $\Theta(b)$,  a function of the impact parameter $b$, 
is  calculated as 

\begin{eqnarray}
\Theta(b) = \pi -2 \phi(b,\infty ).
\end{eqnarray}

where $d$ is the  distance of closest approach for head-on collisions,
\begin{equation}
 d = \frac{ Z_1 Z_2 e^2 }{E},
\end{equation}  
where  particle charges are $Z_1 e$ and $Z_2 e$, the Coulomb interaction
can be written as

\begin{equation}
V_C(r) = \frac{E d}{ r}.
\end{equation}
  
 In this work  the potential $V(r)$ is taken as the sum of the repulsive 
Coulomb potential $V_C(r)$ and an attractive potential $V_N(r)$, given by

\begin{equation}
V(r) = V_C(r) +  V_N(r).
\end{equation}

For the Coulomb potential alone, the distance of closest approach is denoted 
by $a_C$, given by

\begin{eqnarray}
a_C = \frac{d}{2} + \sqrt{\frac{d^2}{4} + b^2 }.
\end{eqnarray}

The Coulomb trajectory is given by
\begin{eqnarray}
\phi_C (b,r)=\int_{a}^{r} dr \frac{ b}{ r^2 \sqrt{1-\frac{d}
 { r}-\frac{b^2}{r^2}}},\\
\mbox{or}~~ \phi_C (b,r)= - W(b,r,0)\Bigg{|}^r _a \\       
\mbox{or}~~ \phi_C (b,r) = \frac {\pi}{2} - W(b,r,0),
\end{eqnarray}
where we define
\begin{equation}
W(b,r,y) = \tan^{-1}\frac{\frac{d}{ 2 b}+\frac{b}{r}}
        { \sqrt{1 + \frac{y}{E}- \frac{d}{r}- \frac{b^2}{r^2}}}.     
\end{equation}
The lower limit, corresponding to $- W(b,a, 0)$, yields $- \frac {\pi }{2}$. 

The deflection function is then given by
\begin{eqnarray}
\Theta_C(b) = 2 W(b, \infty, 0 ), \\
\mbox{or ~~~}  
\Theta_C (b) = 2 \tan^{-1} {\frac{ d}{2 b}}.
\end{eqnarray}

These are standard results available in literature [1-4]. Here the details for
evaluating the integral have been provided, because adding  a single
square well potential or a sum of many square well potentials,
similar integrals arise and can be evaluated.

\section{Trajectories for Coulomb plus  attractive square-well potential}

We apply the ideas of the previous  section by choosing $V_N(r)$ to be
an attractive square well potential  written as

\begin{equation}
V_N(r) =  -V_0 ~U(r - R_0 ),
\end{equation}
of strength $V_0$ and range $R_0$. The function $ U $ is the unit step
function given by 
\begin{eqnarray}
   U( x -x_0 ) = 1,  ~~x \leq x_0 \\
\mbox{and} ~~~ U( x - x_0 ) = 0,  ~~ x > x_0.
\end{eqnarray}

The effective potential consisting of the Coulomb, the attractive  well  and 
the centrifugal potentials can have an  outer barrier maximum at $ r <R_0 $. 
For large impact parameters, the energy $E$ is not sufficient to overcome the
barrier, the distance of closest approach remains $ a_C$.
As the impact parameter is reduced, the centrifugal barrier $\frac{b^2}{r^2}$
decreases, the energy $E$ is above the barrier maximum and $a$ becomes less than
$ R_0$. In this case $a_C$ will be larger than $a$, however it is also less than
$ R_0$ though  $V_N(r)$ is not included in its computation. So the distance of 
closest approach, $a$ can be written as
\begin{equation}
a = a_C U(R_0 -a_C) + a_0 U( a_C -R_0), 
\end{equation}
where
\begin{equation}
a_0 = \frac {\frac{d}{2} + \sqrt{\frac{d^2}{4} + b^2 (1+\frac {V_0}{E}) } }
 {1+ \frac {V_0}{E} }.
\end{equation}

The discontinuity in the potential arises due to a discontinuity in 
the distance of closest approach as a function of impact parameter.

The first term of $a$ occurs when the attractive potential is not seen while the
second term arises when the top of the barrier is overcome. In the latter case, the integral
splits into two, the first one is between $r$ and $R_0$ while the second one is between
$R_0$ and $a$. The expression for the trajectory with  $ r < R_0 $ is  given by 
\begin{equation} 
\phi(b, r)= \frac{\pi}{2} - W(b, r , V_0).
\end{equation}

 The equation for  the trajectory  with $r > R_0$
is given by
\begin{equation} 
\phi(b,r)= \phi_C(b,r) + \phi_0(b,R_0) U(a_C - R_0),
\end{equation}
where $\phi_0$ is given by
\begin{equation}
\phi_0 (b,R_0) =  W(b,R_0, 0) - W(b, R_0, V_0 ). 
\end{equation} 
It is worth noting that $\phi_0$ is not a function of $r$.

The deflection function for this case  is given by

\begin{equation} 
\Theta( b )= \Theta_C (b) - 2 \phi_0 (b, R_0) U( a_C - R_0).
\end{equation}
The first term can be interpreted as the Coulomb term while the second term 
can be interpreted to arise due to the attractive potential and is effective 
only when the Coulomb barrier is overcome. This expression is similar to that
of the quantal scattering amplitude $f(\theta)$  where the Coulomb scattering amplitude 
gets added to the nuclear amplitude in the presence of the Coulomb interaction.

\section{Trajectories for Coulomb plus attractive staircase potential}
 
The  attractive  staircase potential is written as
\begin{equation}
V_N(r) =  -\sum_{i=0} ^m v_i U(r-R_i),
\end{equation}
as a sum of $m + 1$ square wells each of strength $v_i$ and range $R_i$
for $ i = 0 $  to  $m$. We choose $R_m < R_{m-1} < R_{m-2} .....< R_0 $.
In this case there is a local  maximum of the barrier arising at every step and the distance 
of closest approach $a$ as a function of impact parameter has $ m+1 $
discontinuities. These are then reflected in the equations of the trajectory
and the deflection function.
First the expression for $a$, the distance of closest approach is to 
be determined as it is the lower limit of the integral of the equation for
the trajectory. If $a$ lies between $R_s $ and $ R_{s+1}$, we denote it
as $ a_s  $. Let $V_s  = \sum _ {i=0} ^s v_i $, then $a_s$ can be obtained 
by solving the equation 
\begin{eqnarray}
(1 + \frac{V_s}{E}) a_s^2 - d a_s - b^2 =0 \\
\mbox{yielding} ~~~  
a_s = \frac {\frac{d}{2} + \sqrt{\frac{d^2}{4} + b^2 (1+\frac {V_s}{E}) } }
 {1+ \frac {V_s}{E}}.
\end{eqnarray}

To determine $a_s$, first
we calculate all the $a's$ as, $ a_m, a_{m-1}, a_{m-2}, ....,  a_0, a_C$
and compare the ratios
$ \frac{R_m}{a_{m-1}}, \frac{R_{m-1}}{a_{m-2}}, ...., \frac{ R_0}{a_C} $
with one. If the first ratio is greater than one, $a = a_m $.
If it is less than one, the second ratio is compared with one, if it is found 
greater than one, $ a = a_{m-1}$, otherwise the procedure is continued till $a$
is determined as $ a_s$. This implies that the potential between $R_s$
and $R_{s+1}$ becomes effective in defining the trajectory.

The expression of trajectory  for $ a_s < r < R_s $ is given by
\begin{eqnarray} 
\phi (b,r) = \frac{\pi}{2} - W(b,r,  V_s).
\end{eqnarray}
 
The expression of trajectory  for $ R_j <r < R_{j-1} < R_0$ is given by
\begin{eqnarray} 
\phi (b, r) = \frac{\pi}{2} - W(b,r, V_{j-1}) + 
 \sum_{i=j}^s ( W( b, R_i, V_{i-1}) - W(b, R_i, V_i)).
\end{eqnarray}

Defining $V_{-1} =0$, we write the expression of trajectory  for $r > R_0$ as 
 
\begin{equation} 
\phi(b, r)= \phi_C (b, r) + \phi_N (b, R_s) U(a_C - R_0),
\end{equation}
where
\begin{eqnarray} 
\phi_N (b, R_s) = \sum_{i=0}^s (W(b, R_i, V_{i-1}) - W(b, R_i, V_i)).
\end{eqnarray} 

In the expression of $\phi (b, r)$,  the first  term can be considered to  
arise in pure 
Coulomb field while the second term  represents the effect of  the attractive potential. 
                                                      
The deflection function, $\Theta(b)$ is then calculated by taking $ r$ going 
to $\infty$ as 
\begin{eqnarray}
\Theta(b) = \Theta_C(b) - 2 \phi_N (b, R_s) U (a_C  - R_0).
\end{eqnarray}

\section{Conclusions}
If usually encountered 
nuclear potential  of the Woods-Saxon form  is approximated by a few discrete 
steps,  analytical expressions of the present paper for trajectories and deflection 
function can be used to gain more insight into the scattering processes. 
Expressions for  trajectories  in the case of repulsive Coulomb plus attractive
multistep potentials can also be utilised 
for many classical and semiclassical formulations in heavy-ion nuclear physics
to describe scattering  and other 
phenomena such as fusion and deep inelastic  scattering. 

Trajectories and deflection function for any potential, $V(r)$ which can be 
parametrized as $ V(r) = \alpha -\frac{\beta}{r}- \frac{\gamma}{r^2}$ piecewise
can be calculated by the method described in this paper. The present formulation
holds even when the Coulomb potential is absent. 
\vspace{1cm}

{\bf Acknowledgments}\\

We thank Sudhir Jain for going through the manuscript and for making useful
comments.
\noindent

{\bf References\\}

[1] K.W. Ford and J.A. Wheeler, Ann. Phys. 7(1959)287.

[2] N.F. Mott and N.S.W. Massey, The Theory of Atomic Collisions\\
published by Oxford University Press, (1965, 3rd edition) p. 97-110.

[3] R.A. Broglia and A. Winther, Heavy Ion Reactions, Vol. I\\
published by Addison-Wesley Publishing Co. (1996) p. 174-180.

[4] P. Froebrich and R. Lipperheide, Theory of Nuclear Reactions\\ 
published by Clarendon Press, Oxford (1996) p.34.

\end{document}